\documentclass{article}
\usepackage{amssymb}

\usepackage{sw20ssur}

\input{tcilatex}

\begin{document}

\author{Azhar Iqbal \\
Department of Mathematics, University of Hull, HU6 7RX, UK}
\title{Impact of entanglement on the game-theoretical concept of evolutionary
stability}
\date{August 20, 2005}
\maketitle

\begin{abstract}
An example of a quantum game is presented that explicitly shows the impact
of entanglement on the game-theoretical concept of evolutionary stability.
\end{abstract}



\section{Introduction}

One of the major motivations in recent work on quantum games \cite
{Mermin,Vaidman,MeyerDavid,Eisert,Eisert1,Enk,Benjamin,Benjamin1,Marinatto,Du,Johnson,Piotrowski,Flitney,Cheon,Iqbal}
is to know how quantization of a game affects/changes the existence/location
of a Nash equilibrium (NE) \cite{Leonard}. Game theory \cite{Neumann} offers
examples where, instead of a unique equilibrium, multiple Nash equilibria
emerge as the solutions of a game. In these examples selecting one (or
possibly more) from many equilibria requires a refinement of the NE concept.
A refinement is a rule/criterion that prefers some equilibria out of many.
Numerous refinements are found in game theory; examples include perfect
equilibrium (used for extensive- and normal-form games), sequential
equilibrium (a fundamental non-cooperative solution concept for
extensive-form games) and correlated equilibrium (used for modelling
communication among players).

From the view point of quantum games the notion of refinement of NE
naturally motivates the question how quantization of a game can affect a
refinement; without affecting the corresponding NE. That is, a NE persists%
\footnote{%
For two-player games by saying that a NE persists in both the classical and
quantum version of a game means that there exists a NE consisting of quantum
strategies that rewards both the players exactly the same as does the
corresponding NE in the classical version of the game.} in both the
classical and quantum versions of a game but its refinement does not. Games
and quantization procedures offering such examples can be said to extend the
boundary of investigations in quantum games from existence/location of NE to
existence/location of one (or more) of its refinements -- in relation to
quantization of the game. Present paper offers such an example.

The rest of the paper is organized as follows. Section ($2$) discusses the
concept of evolutionary stability and why is can be discussed in quantum
games. In Section ($3$) an example is presented that shows explicitly the
role of entanglement in deciding evolutionary stability of a strategy. For
comparison a separate subsection discusses the classical case. Section ($4$)
discusses results and indicates lines of further investigation.

\section{Evolutionary stability}

The concept of an Evolutionarily Stable Strategy (ESS) \cite{Maynard
Smith,Maynard Smith1} is the cornerstone of evolutionary game theory \cite
{Hofbauer}; introduced some thirty years ago by mathematical biologists as a
game-theoretical model to understand equilibrium behavior of a system of
interacting entities. Their original intention was to explain phenomena like
polymorphism\footnote{%
Polymorphism is the occurrence of different forms, stages or types in
individual organisms or in organisms of the same species.} of behavior in
animal societies. These phenomena happen in spite of the fact that the
individuals in these societies have neither conscience; nor rationality; nor
expectations; nor even the choice between several behavioral patterns which
are thought to be determined genetically. Sometimes, even using the term
``individual'' for members of these animals has itself been questioned.

Roughly speaking, ESS is described as a strategy having the property that if
all members of a population adapt it, no mutant strategy could invade the
population under the influence of natural selection. That is, if a strategy
is adapted by at least $1-\epsilon _{0}$ fraction of the population then it
can resist invasion by any $\epsilon <\epsilon _{0}$ mutant strategies. ESS
is generally accepted as another refinement of the NE concept; but the
concept has been given mathematical formulations in several contexts \cite
{Lessard}. In the familiar context, ESS is described as a refinement on the
set of symmetric Nash equilibria which is robust against small changes
(mutations) that may appear in prevalent strategy. The robustness against
mutations is referred to as \textit{evolutionary stability}.

The definition of an ESS, as it was originally introduced \cite{Maynard
Smith} in mathematical biology, assumes \cite{Bomze}:

\begin{itemize}
\item  An infinite population of players engaged in random pair-wise
contests.

\item  Each player being programmed to play only one strategy.

\item  An evolutionary pressure ensuring that the better players have better
chances of survival.
\end{itemize}

It is seen that the setting of evolutionary game theory diverges away from
the usual setting of game theory in disassociating its players from the
capacity to make rational decisions. The players' strategies are inheritable
traits (phenotypes) that evolution tests for their suitability and value in
the players' struggle for survival over the course of time. In contrast, the
usual approach in game theory assumes players both capable of making
rational decisions and always interested to maximize their payoffs.

Interestingly, in parallel to its original formulation, the ESS theory can
also be interpreted such that a player's strategy is not a phenotype but an
available option or a possible state attributable to a player. With this
interpretation the \emph{same} players continue their repeated pair-wise
contests. The evolutionary pressure, however, now ensures the survival of
better-performing strategies, \emph{not} players. ESS will emerge as a
strategy which if played by most of the population will withstand invasion
from mutant strategies that are played by small number of the members of the
population. This view of the ESS concept can be studied also in quantum
mechanical versions of evolutionary games because it allows replacement of
classical strategies with their quantum analogues. Moreover, the view
permits not to associate with the players a capacity to make rational
decisions; thus retaining intact one of the important features of the ESS
theory.

A question can be raised here: Where in nature do the quantum strategies
exist and are being subjected to evolutionary pressures? It seems that
example of inter-molecular interactions can be promising candidate. These
interactions can both be pair-wise and taking place under evolutionary
forces. For these interactions the players' disassociation from their
capacity to act rationally can also be granted, without further assumptions.

We now move to a question of interest that such setting is bound to raise:
How game-theoretical solution concepts, which are especially developed for
the understanding of evolutionary dynamics, adapt/shape/change themselves
with players having access to quantum strategies? For the possible situation
of inter-molecular interactions this question shapes itself to ask how
game-theoretical solution concepts, that are especially developed for
evolutionary games, predict different equilibrium states of a population
consisting of interacting molecules to which quantum strategies can be
associated.

The emerging field of quantum games recognizes entanglement as a resource
giving new, and often counter-intuitive dimensions to world of playing
games. From the view point of ESS theory the players' sharing of a new
resource of entanglement leads one to ask whether entanglement can change
the evolutionary stability of a symmetric NE. Of course, during this change
the corresponding NE remains intact in both the classical and quantum forms
of the game. Evolutionary stability is a solution concept having a relevance
to a population as a whole with regards to its capacity to withstand mutant
strategies. If entanglement is found to decide evolutionary stability then
the lesson from the ESS theory is that entanglement's effects are not
confined to the pair-wise encounters but the phenomenon can also decide the
fate of the whole population, in terms of its susceptibility to invasion
from the mutant strategies that appear in small numbers.

\section{An example}

Though recent work in quantum games presents examples \cite{Iqbal} of games
where evolutionary stability is related to quantization of a game, a direct
and explicit relationship between a measure of entanglement and the
mathematical concept of evolutionary stability is still to be investigated
even for two-player games. Earlier work \cite{Iqbal} on this topic uses a
particular quantization scheme for matrix games suggested in the Ref. \cite
{Marinatto}. In this scheme a measure of entanglement does not appear
explicitly in players' payoff expressions. In Eisert et al.'s scheme, on the
other hand, players' payoffs contain entanglement explicitly, which makes
possible, in the present contribution, to develope an example that shows the
relationship between entanglement and evolutionary stability.

Consider a symmetric bi-matrix game given by the matrix:

\begin{equation}
\begin{array}{c}
\text{Alice}
\end{array}
\begin{array}{c}
S_{1} \\ 
S_{2}
\end{array}
\stackrel{\stackrel{
\begin{array}{c}
\text{Bob}
\end{array}
}{
\begin{array}{cc}
S_{1} & S_{2}
\end{array}
}}{\left( 
\begin{array}{cc}
(r,r) & (s,t) \\ 
(t,s) & (u,u)
\end{array}
\right) }  \label{Matrix}
\end{equation}
Suppose Alice and Bob play the strategy $S_{1}$ with probabilities $p$ and $%
q $, respectively. The strategy $S_{2}$ is then played with probabilities $%
(1-p)$ and $(1-q)$ by Alice and Bob, respectively. We denote Alice's payoff
by $P_{A}(p,q)$ when she plays $p$ and Bob plays $q$. That is, Alice's and
Bob's strategies are identified from the numbers $p,q\in \lbrack 0,1]$,
without referring to $S_{1}$ and $S_{2}$. For the matrix (\ref{Matrix}) the
Alice's payoff $P_{A}(p,q)$, for example, reads

\begin{equation}
P_{A}(p,q)=rpq+sp(1-q)+t(1-p)q+u(1-p)(1-q)  \label{Payoffs}
\end{equation}
Similarly, Bob's payoff $P_{B}(p,q)$ can be written. In this symmetric game
we have $P_{A}(p,q)=P_{B}(q,p)$ and, without using subscripts, $P(p,q)$, for
example, describes the payoff to $p$-player against $q$-player. In this game
the inequality

\begin{equation}
P(p^{\ast },p^{\ast })-P(p,p^{\ast })\geqslant 0  \label{NE}
\end{equation}
says that the strategy $p^{\ast }$, played by both the players, is a NE.

\subsection{Evolutionary stability: classical game}

For symmetric contests an Evolutionarily Stable Strategy (ESS) is defined as
follows. A strategy $p^{\ast }$ is an ESS when

\begin{equation}
P(p^{\ast },p^{\ast })>P(p,p^{\ast })  \label{ESSFirst}
\end{equation}
that is, $p^{\ast }$ is a strict NE. If it is not the case and

\begin{equation}
P(p^{\ast },p^{\ast })=P(p,p^{\ast })\text{ then }p^{\ast }\text{ is ESS if }%
P(p^{\ast },p)>P(p,p)  \label{ESSSecond}
\end{equation}
showing that every ESS is a NE but not otherwise.

We consider a case when

\begin{equation}
s=t,\text{ \ \ }r=u\text{ and \ \ }(r-t)>0  \label{GameDefinition}
\end{equation}
in the matrix (\ref{Matrix}). In this case the inequality (\ref{NE}) along
with the definition (\ref{Payoffs}) gives

\begin{equation}
P(p^{\ast },p^{\ast })-P(p,p^{\ast })=(p^{\ast }-p)(r-t)(2p^{\ast }-1)
\label{PayoffDiff}
\end{equation}
and the strategy $p^{\ast }=1/2$ comes out as a mixed NE. From the defining
inequalities (\ref{ESSFirst},\ref{ESSSecond}) we get $P(1/2,1/2)-P(p,1/2)=0$
and the first condition (\ref{ESSFirst}) of an ESS does not apply. The
second condition (\ref{ESSSecond}), then, gives

\begin{equation}
P(1/2,p)-P(p,p)=(r-t)\left\{ 2p(1-p)-1/2\right\}  \label{PayoffDiff1}
\end{equation}
which can not be strictly greater than zero given $(r-t)>0$. For example, at 
$p=0$ it becomes a negative quantity. Therefore, for the matrix game defined
by (\ref{Matrix}) and (\ref{GameDefinition}) the strategy $p^{\ast }=1/2$ is
a symmetric NE, but it is not evolutionarily stable. Also, at this
equilibrium both players get $(r+t)/2$ as their payoffs.

\subsection{\label{ESSQuantumGame}Evolutionary stability: quantum game}

Consider the \emph{same} game, i.e. defined by (\ref{Matrix}) and (\ref
{GameDefinition}), played now by the set-up proposed by Eisert et al. \cite
{Eisert,Eisert1}. This scheme suggests a quantum version of the game (\ref
{Matrix}) by assigning two basis vectors $\left| S_{1}\right\rangle $ and $%
\left| S_{2}\right\rangle $ in the Hilbert space of a qubit. States of the
two qubits belong to two-dimensional Hilbert spaces $H_{A}$ and $H_{B}$
respectively. State of the game is defined by a vector residing in the
tensor-product space $H_{A}\otimes H_{B}$, spanned by the basis $\left|
S_{1}S_{1}\right\rangle ,\left| S_{1}S_{2}\right\rangle ,\left|
S_{2}S_{1}\right\rangle $ and $\left| S_{2}S_{2}\right\rangle $. Game's
initial state is $\left| \psi _{ini}\right\rangle =\hat{J}\left|
S_{1}S_{1}\right\rangle $ where $\hat{J}$ is a unitary operator known to
both the players. Alice's and Bob's strategies are unitary operations $\hat{U%
}_{A}$ and $\hat{U}_{B}$, respectively, chosen from a strategic space \c{S}.
The state of the game changes to $\hat{U}_{A}\otimes \hat{U}_{B}\hat{J}%
\left| S_{1}S_{1}\right\rangle $ after players' actions. Finally,
measurement consists of applying reverse unitary operator $\hat{J}^{\dagger
} $ followed by a pair of Stern-Gerlach type detectors. Before detection the
final state of the game is $\left| \psi _{fin}\right\rangle =\hat{J}%
^{\dagger }\hat{U}_{A}\otimes \hat{U}_{B}\hat{J}\left|
S_{1}S_{1}\right\rangle $. The players' expected payoffs can then be written
as the projections of the state $\left| \psi _{fin}\right\rangle $ onto the
basis vectors of tensor-product space $H_{A}\otimes H_{B}$, weighed by the
constants appearing in the game (\ref{Matrix}). For example, Alice's payoff
reads

\begin{equation}
P_{A}=r\left| \left\langle S_{1}S_{1}\mid \psi _{fin}\right\rangle \right|
^{2}+s\left| \left\langle S_{1}S_{2}\mid \psi _{fin}\right\rangle \right|
^{2}+t\left| \left\langle S_{2}S_{1}\mid \psi _{fin}\right\rangle \right|
^{2}+u\left| \left\langle S_{2}S_{2}\mid \psi _{fin}\right\rangle \right|
^{2}  \label{Eisert's Alice's payoff}
\end{equation}
Bob's payoff is, then, obtained by the transformation $s\rightleftarrows t$
in Eq. (\ref{Eisert's Alice's payoff}). Eisert et al. \cite{Eisert,Eisert1}
allowed players' actions from the space \c{S} of unitary operators of the
form

\begin{equation}
U(\theta ,\phi )=\left( 
\begin{tabular}{ll}
e$^{i\phi }\cos (\theta /2)$ & $\sin (\theta /2)$ \\ 
$\text{-}\sin (\theta /2)$ & e$^{-i\phi }\cos (\theta /2)$%
\end{tabular}
\right)  \label{Eisert's unitary operators}
\end{equation}
where

\begin{equation}
\theta \in \lbrack 0,\pi ]\text{ and }\phi \in \lbrack 0,\pi /2]
\label{Ranges}
\end{equation}
They defined their unitary operator $\hat{J}=\exp \left\{ i\gamma
S_{2}\otimes S_{2}/2\right\} $ with $\gamma \in \lbrack 0,\pi /2]$
representing a measure of the game's entanglement. At $\gamma =0$ the game
reduces to its classical form.

After this note on Eiset et al.'s scheme, we set $s_{A}\equiv (\theta
_{A},\phi _{A})$ and $s_{B}\equiv (\theta _{B},\phi _{B})$ to denote Alice's
and Bob's strategies, respectively. Because the quantum game is symmetric
i.e. $P_{A}(s_{A},s_{B})=P_{B}(s_{B},s_{A})$ we can write, as before, $%
P(s_{A},s_{B})$ for the payoff to $s_{A}$-player against $s_{B}$-player. For
quantum form of the game (\ref{Matrix},\ref{GameDefinition}) one finds

\begin{equation}
P(s_{A},s_{B})=(1/2)(r-t)\left\{ 1+\cos \theta _{A}\cos \theta _{B}+\sin
\theta _{A}\sin \theta _{B}\sin \gamma \sin (\phi _{A}+\phi _{B})\right\} +t
\label{PayoffsQG}
\end{equation}
The definition of a NE gives $P(s^{\ast },s^{\ast })-P(s,s^{\ast })\geqslant
0$ where $s=(\theta ,\phi )$ and $s^{\ast }=(\theta ^{\ast },\phi ^{\ast })$%
. The definition can be written as

\begin{equation}
\left\{ \partial _{\theta }P\mid _{\theta ^{\ast },\phi ^{\ast }}(\theta
^{\ast }-\theta )+\partial _{\phi }P\mid _{\theta ^{\ast },\phi ^{\ast
}}(\phi ^{\ast }-\phi )\right\} \geq 0
\end{equation}
We search for a quantum strategy $s^{\ast }=(\theta ^{\ast },\phi ^{\ast })$
for which both $\partial _{\theta }P\mid _{\theta ^{\ast },\phi ^{\ast
}},\partial _{\phi }P\mid _{\theta ^{\ast },\phi ^{\ast }}$ vanish at $%
\gamma =0$ and at some other value of $\gamma $ which is not zero. For the
payoffs (\ref{PayoffsQG}) the strategy $s^{\ast }=(\pi /2,\pi /4)$ satisfies
these conditions. For this strategy the Eq. (\ref{PayoffsQG}) gives

\begin{equation}
P(s^{\ast },s^{\ast })-P(s,s^{\ast })=(1/2)(r-t)\sin \gamma \left\{ 1-\sin
(\phi +\pi /4)\sin \theta \right\}  \label{NEQG}
\end{equation}
At $\gamma =0$ the strategy $s^{\ast }=(\pi /2,\pi /4)$, when played by both
the players, is a NE which rewards the players same as does the strategy $%
p^{\ast }=1/2$ in the classical version of the game i.e. $(r+t)/2$. Also,
then we have $P(s^{\ast },s^{\ast })-P(s,s^{\ast })=0$ from Eq. (\ref{NEQG})
and the ESS's second condition (\ref{ESSSecond}) applies. Use Eq. (\ref
{PayoffsQG}) to evaluate

\begin{equation}
P(s^{\ast },s)-P(s,s)=-(r-t)\cos ^{2}(\theta )+(1/2)(r-t)\sin \gamma \sin
\theta \left\{ \sin (\phi +\pi /4)-\sin \theta \sin (2\phi )\right\}
\label{ESSSecondQ}
\end{equation}
which at $\gamma =0$ reduces to $P(s^{\ast },s)-P(s,s)=-(r-t)\cos
^{2}(\theta )$, which can assume negative values. The game's definition (\ref
{GameDefinition}) and the ESS's second condition (\ref{ESSSecond}) show that
the strategy $s^{\ast }=(\pi /2,\pi /4)$ is not evolutionarily stable at $%
\gamma =0$.

Now consider the case when $\gamma \neq 0$ to know about the evolutionary
stability of the \emph{same} quantum strategy. From (\ref{Ranges}) we have
both $\sin \theta ,\sin (\phi +\pi /4)\in \lbrack 0,1]$ and the Eq. (\ref
{NEQG}) indicates that $s^{\ast }=(\pi /2,\pi /4)$ remains a NE for all $%
\gamma \in \lbrack 0,\pi /2]$. The product $\sin (\phi +\pi /4)\sin \theta $
attains a value of $1$ only at $s^{\ast }=(\pi /2,\pi /4)$ and remains less
than $1$ otherwise. The Eq. (\ref{NEQG}) shows that for $\gamma \neq 0$ the
strategy $s^{\ast }=(\pi /2,\pi /4)$ becomes a strict NE for which the ESS's
first condition (\ref{ESSFirst}) applies. Therefore, for the game defined in
(\ref{GameDefinition}) the strategy $s^{\ast }=(\pi /2,\pi /4)$ is
evolutionarily stable for a non-zero measure of entanglement $\gamma $.

\section{Discussion}

The above example shows explicitly how presence of entanglement leads to
evolutionary stability of a strategy. It is of interest for three apparent
reasons. Firstly, the game-theoretical concept of evolutionary stability has
very rich literature in game theory, mathematical biology and evolutionary
economics \cite{Friedman,Evolutionary economics}. Secondly, the result that
the game-theoretical concept of evolutionary stability can be discussed in
relation to entanglement opens a new role for this phenomenon. It is a role
where entanglement decides whether a population of interacting entities can
withstand invasion from mutant strategies appearing in small numbers.
Thirdly, this extended role for entanglement can possibly be helpful to
better understand entanglement itself.

A possible criticism of studying evolutionary stability in quantum games may
come from the following view point. Being a game-theoretical solution
concept, originally developed to understand problems in population biology,
how can the concept be taken out of its context of population biology and
discussed in quantum games? Evolutionary stability was indeed originally
introduced within population biology but the concept can also be given an
interpretation in term of two-player game that is infinitely repeated.
Secondly, the population setting that evolutionary stability assumes does
not come only from discussion of the problems of population biology. Even
the concept of NE, as it was originally developed by John Nash, assumed a
population of players. In his unpublished thesis he wrote `\textit{it is
unnecessary to assume that the participants have...... the ability to go
through any complex reasoning process. But the participants are supposed to
accumulate empirical information on the various pure strategies at their
disposal.......We assume that there is a population .......of
participants......and that there is a stable average frequency with which a
pure strategy is employed by the ``average member'' of the appropriate
population}'\cite{Hofbauer,Leonard}. Evolutionary stability, as a
game-theoretical concept, also has roots in efforts to get the game theory
rid of its usual approach devoting itself to analyzing games among
hyper-rational players always ready and engaged in selfish interests to
increase their payoffs. The lesson evolutionary stability teaches is that
playing games can be disassociated from players' capacity to make rational
decisions. Such disassociation seems equally fruitful to possible situations
where quantum games are being played in nature; because associating
rationality to quantum interacting entities is of even much more remote
possibility then bacteria and viruses whose behavior evolutionary game
theory explains.

An interesting approach \cite{Entropy} characterizes ESSs in terms of
extremal states of a function known as \textit{evolutionary entropy} defined
by

\[
E=-\stackunder{i}{\sum }\mu _{i}\log \mu _{i} 
\]
where $\mu _{i}$ represents the relative contribution of the $i$th strategy
to the total payoff. A possible extension of the present approach may be the
case when entanglement decides extremal states of evolutionary entropy.
Extension on similar lines can be proposed for another quantity which Bomze
called \textit{relative negentropy} \cite{Bomze} and it is optimized during
the course of evolution.

In the Section (\ref{ESSQuantumGame}) entanglement gives evolutionary
stability to a symmetric NE by making it a strict NE as well. Thus
evolutionary stability is achieved by only using the ESS's first condition.
Perhaps a more interesting example is when entanglement gives evolutionary
stability via the ESS's second condition. That is, entanglement makes $%
P(s^{\ast },s)$ strictly greater than $P(s,s)$ when $P(s^{\ast },s^{\ast })$
and $P(s,s^{\ast })$ are equal.

\end{document}